\documentclass[12pt,preprint]{aastex}

\newcommand{\as}{''}
\newcommand{\simgt}{\lower.5ex\hbox{$\; \buildrel > \over \sim \;$}}
\newcommand{\simlt}{\lower.5ex\hbox{$\; \buildrel < \over \sim \;$}}
\newcommand{\sol}{_\odot}
\newcommand{\kms}{km s$^{-1}$}

\begin{document}
\title{How Dry Are Red Mergers?}
\author{Jennifer L. Donovan\altaffilmark{1}, J.~E. Hibbard\altaffilmark{2}, J.~H. van Gorkom\altaffilmark{1}}
\altaffiltext{1}{Department of Astronomy, Columbia University, 550 West 120th St., Mail Code 5246, New York, NY 10027; jen@astro.columbia.edu, jvangork@astro.columbia.edu}
\altaffiltext{2}{NRAO; jhibbard@nrao.edu}

\begin{abstract}

The focus of current research in galaxy evolution has increasingly turned to understanding the effect that mergers have on the evolution of systems on the red sequence. For those interactions purported to occur dissipationlessly (so called ``dry mergers"), it would appear that the role of gas is minimal. However, if these mergers are not completely dry, then even low levels of gas may be detectable. The purpose of our study is to test whether early type galaxies with HI in or around them, or ``wet" ellipticals, would have been selected as dry mergers by the criteria in van Dokkum (2005, AJ, 130, 2647). To that end, we examine a sample of 20 early types from the HI Rogues Gallery with neutral hydrogen in their immediate environs. Of these, the 15 brightest and reddest galaxies match the optical dry merger criteria, but in each case, the presence of HI (for the majority, at levels $>$10$^{8}$ M$\sol$) -- as well as significant star formation in some cases -- means that they are not truly dry. 

\end{abstract}


\section{Introduction}

The ways in which early type galaxies form and evolve continue to be matters of much debate. Proposed mechanisms of formation have gone full circle, from monolithic collapse (e.g. \citealt{Eggen62}) through late mergers of fully formed disk galaxies (e.g. \citealt{Toomre72}, \citealt{Galaxies96}), back to very early formation of at least the stars (though perhaps not the assembly of the galaxies themselves, e.g. \citealt{Bower92}, \citealt{Ellis97}, \citealt{vanDokkum98}.) Certainly observations suggest that at least some ellipticals were formed via mergers of gas-rich disks (e.g. \citealt{Toomre72}, \citealt{Schweizer82}, \citealt{Hibbard96}), though it is not obvious how common this formation history is. More recently, the debate has focussed on the role of dissipationless merging between gas-poor galaxies (e.g. \citealt{Ciotti06}, \citealt{Scarlata07}). \\
\indent If mergers are indeed central to the process of early type galaxy formation, then the nature of the progenitors and the characteristics of the encounter become essential in describing the evolution of these systems. The only clues available for discerning the evolutionary history of a galaxy are the nature of its composite populations. Using COMBO-17 survey data, \citet{Bell04} found that the models which best fit the evolution of the mean color of 4690 red sequence galaxies between $0.2<z\leq1.1$ (the vast majority of which are early types) are those of passively aging stellar populations. This is consistent with the majority of the galaxies' stars (if not the galaxies themselves) being formed at redshifts $>$2. Recent studies, however, have also produced observations consistent with a dusting of more recent star formation in some ellipticals with otherwise ancient stars (e.g. \citealt{Trager00}). \\
\indent Studying the B-band luminosity density evolution with redshift of the above early-type galaxies over the range of $0.2<z\leq1.1$ reveals another clue about the histories of these systems; the stellar mass on the red sequence increases by a factor of 2-3 over this time \citep{Bell04}. Such an increase in mass without significant star formation argues in favor of either very dusty or dissipationless mergers. Other questions have remained unanswered by the dissipational merger theory. For instance, it is surprising that mergers of spirals containing a range of stellar populations would produce the observed low scatter in the fundamental plane (e.g. \citealt{Bower98}).\\
\indent Arguments in favor of dissipational mergers, on the other hand, include the observed signatures that these interactions involving disks leave behind in the form of fine structures. Shells and ripples are observed in a significant fraction of nearby ellipticals (\citealt{Malin83}, \citealt{Schweizer90}), indicating the presence of a dynamically cold component, e.g. a disk \citep{Hernquist88}, in the progenitors. In most disk galaxies, particularly later Hubble types, such disks are gas rich \citep{Roberts94}. \\
\indent Recently, \citet[hereafter VD05]{vanDokkum05} studied the 126 brightest, reddest field galaxies in the NOAO Deep Wide Field Survey (NDWFS) and the Multi-wavelength Survey by Yale-Chile (MUSYC). He found that of this sample, 71\% of the 86 systems defined as E/S0 exhibit tidal features, such as tails or broad fans of stars, which are attributed to mergers or interactions. This is taken as evidence by van Dokkum for red or ``dry" mergers, wherein red, bulge-dominated galaxies experience nearly dissipationless, or gasless, mergers in order to form massive ellipticals. Using only the ongoing mergers within this subsample, VD05 derived a mass accretion rate for galaxies on the red sequence of $\Delta M/M$ = 0.09 $\pm$ 0.04 Gyr$^{-1}$. From this rate, it can be inferred that between $0<z<1$, the masses of bright red galaxies have undergone an increase in their stellar mass density by a factor of $\geq$2 provided this rate stays constant or increases with redshift, consistent with the COMBO-17 data presented in \citet{Bell04}. Thus, VD05 argues that dry mergers at low redshift are responsible for much of the local bright field elliptical galaxy population. However, no information regarding the gas content of these systems was presented. \\ 
\indent Also from the NDWFS, \citet{Brown07} compiled a sample of red galaxies to z=1 that is an order of magnitude larger than similar surveys in the literature. These authors examine the B-band luminosity density evolution of red galaxies and also find that the stellar mass in these systems has increased by a factor of $\sim$2; however, they contend that declining star formation rates in blue galaxies are sufficient to produce L* and smaller red galaxies. On the high mass end of the red sequence, they find that $\approx$80$\%$ of the stellar populations within the largest ($\>4\it{L}$*) red galaxies were in place by z=0.7, and more recent mergers of red galaxies are not the sole cause of the subsequent increase in stellar mass. \citet{Scarlata07} argue against dry mergers as well, indicating that blue, irregular, and disk galaxies can account for the mass increase since z=0.7 in $<2.5\it{L}$* early type galaxies via their evolution (and possibly via dissipational merging) into low- and intermediate-mass red early types. \\ 
\indent At lower redshift, \citet{Schweizer90} and \citet{Malin83} find that a significant fraction of nearby field ellipticals exhibit signs of fine structure such as shells, indicating the presence of a cold component. Among these samples, \citet{vanGorkom97} find that 50\% of ellipticals in gas-rich environments have associated HI at the $\sim$10$^{9}$ M$\sol$ level. So how ``dry" exactly are the mergers in the more complete sample of VD05? What role do mergers of any type play in the evolution of red galaxies on the red sequence? \\
\indent We explore these questions in this paper. Specifically, we select red normal and peculiar galaxies which are morphologically similar to the sample in VD05 and whose HI content we know in order to examine the likelihood that the VD05 sample can be truly ``dry". In \S 2, we discuss the sample of VD05 at z=0.1 and our local analogous sample, and we discuss the results of this comparison in \S 3, followed by our conclusions in \S 4. Throughout the paper, we assume H$_{o}$=75 \kms Mpc$^{-1}$. 
\\

\section{The Samples}

\subsection{NDWFS and MUSYC}
In VD05, the 126 reddest and brightest field galaxies in the NOAO Deep Wide Field Survey (NDWFS) and the Multiwavelength Survey by Yale-Chile (MUSYC) around z=0.1 were selected in order to uniformly study the morphologies of red galaxies between L* and 3L*. The VD05 color selection criteria corresponds to 1.6 $\leq$ (B-R) $\leq$ 2.2, and (B-R) $>$ 1.6 + 0.12(R - 15), and the brightness criterion requires galaxies to have an apparent Cousins R magnitude (hereafter, simply ``R") $>$ 17 (c.f. Figure 2 in VD05). The galaxies were classified according to their optical morphology; approximately half of the sample exhibit low surface brightness features indicative of interactions, such as plumes and tails (VD05). More dramatically, 70\% of the bulge-dominated early-type galaxies show such features; these bright, red, and morphologically selected systems are called red or ``dry mergers" by VD05. The presence of tidal features was revealed by the deep imaging survey data, and each object was further classified as a ``weak," ``strong," or ``ongoing" interaction. The color-magnitude diagram (CMD) of the VD05 sample is shown in Figure 1 with dotted lines highlighting the color and magnitude selection criteria. \\
\indent To make the sample of VD05 consistent with ours for the purpose of comparison, we must adjust it to z=0 and apply a k-correction to both the R magnitudes and B-R colors to account for this change in redshift. The 126 galaxies of VD05 have a median redshift of 0.1, and we know the precise redshifts for $\sim$80$\%$ of this sample; for the other $\sim$20$\%$, we assume z = 0.1. We apply the k-corrections published in \citet{Poggianti97}, utilizing the E model in order to adjust the magnitudes and colors of the 86 galaxies classified as E/S0 as well as the 30 galaxies classified as S0 by VD05; we apply the Sa model to the 10 galaxies classified as S by VD05. The Sa model was not computed for R$_{c}$ magnitudes in \citet{Poggianti97}; for these systems we used the values published for Johnson R magnitudes. The difference between the R and R$_{c}$ k-corrections is less than 0.03 mag out to z=0.24, which is accurate enough for our purposes. \\
\indent Taking these distances and k-corrections into account, the corresponding z=0 CMD of the VD05 sample is shown beside its z=0.1 counterpart in Figure 1, again with the (z=0.1) color and magnitude cutoffs displayed as dotted lines. The galaxies are no longer well constrained within an apparent magnitude-dependent parameter space; nine systems are actually fainter than the magnitude cutoff. For this reason, we define a new magnitude limit of M$_{R}$=-20.5, and we also extend the magnitude-color criterion, (B-R) $>$ 1.6 + 0.12(R - 15), to fainter magnitudes as it still defines well the ``blue" edge of the sample. These new criteria, which encompass all of the VD05 galaxies, are shown by the dashed lines in the right panel of Figure 1. 
Galaxies for which the redshift is unknown (and z=0.1 is assumed) are represented by diamonds in Figure 1. 
\subsection{HI Rogues}
\indent The purpose of our sample is to test whether early types with HI in or around them, or ``wet" ellipticals, would have been selected as dry mergers by the criteria used in VD05. For this purpose, it is useful to study systems listed in the HI Rogues Gallery \citep{RoguesGallery}, a collection of HI images of peculiar galaxies as well as otherwise normal galaxies with peculiar HI morphology. Specifically, we consider those listed as being early types (Rogues classes pecEo, pecEi, EpecH -- peculiar ellipticals with HI outside/inside the optical body and normal ellipticals with peculiar HI), and we include two well known merger remnants for comparison (NGC 3921 and NGC 7252). Our sample consists of galaxies from these classes with B and R magnitudes available in the literature that were measured inside appropriate aperture sizes. The sample and each galaxy's Rogues classification are listed in Table 1. The total number of galaxies in our sample is 20. \\
\subsection{Optical Photometry}
\indent For the 126 galaxies in the VD05 sample, apparent R magnitudes and B-R colors were measured inside an aperture of 5$\as$ diameter. 
The central 5$\as$ diameter area of a galaxy at z=0.1 corresponds to its central 9 kpc diameter area. \\ 
\indent In order to directly compare the HI Rogues selected above to VD05, absolute R magnitudes (M$_{R}$) and B-R colors are determined for each system from existing photometric data in the literature (see Table 1), using the HyperLeda 
database compilation of aperture photometry \citep{Paturel03}. Taking into account the largest quoted uncertainty of these individual studies, the photometry of our sample is accurate to better than 0.1 mag. \\
\indent The aperture sizes necessitated by each particular system in our sample differ since the galaxies are all located at different distances; the aperture size is chosen to match the central 9 kpc of each galaxy as closely as possible in accordance with VD05. In Table 1, we list the adopted distances for each system; distances are taken from the Nearby Galaxy Catalog \citep{Tully88} unless otherwise noted. For the galaxies not listed in \citet{Tully88}, we calculate Hubble flow distances using Virgo-corrected velocities listed in NED. We also list a parameter R$_{ap}$, defined as the ratio of the equivalent linear diameter of the adopted aperture (D$_{ap}$) to 9 kpc, or

\begin{equation}
R_{ap} = \frac{D_{ap}}{9~\rm{kpc}},
\end{equation}

\noindent for each galaxy. When R$_{ap}$ $>$ 1.00, our photometry is from a region larger than 9 kpc in diameter. For most of the sample, R$_{ap}$ is within 10\% of 1.00; the four significant deviations are for NGC 7626 (R$_{ap}$=0.87), IC 2006 and NGC 2865 (R$_{ap}$=0.84), and NGC 3921 (R$_{ap}$=1.26). For the two most distant systems (d $>$ 80 Mpc), NGC 3921 and Mrk 315, we also apply magnitude and color k-corrections as described above. \\ 
\indent We use the magnitude and color criteria specified in \S{2.1} to identify galaxies from our Rogues sample that would have been selected by VD05. The Rogues sample is overplotted on the VD05 sample in Figure 2, where we show the galaxies' absolute magnitudes as a function of B-R color. The VD05 z=0.1 color and apparent magnitude-dependent selection criteria are shown as in Figure 1, and the color-magnitude extension as well as the fainter magnitude cutoff that we use are also displayed. Though the HI sample is overall fainter than the sample of VD05, the overlap between the samples is obvious. \\

\section{Results and Discussion}
\subsection{``Wet" Red Rogues}

\indent As is evident in Figure 2, 15 (75$\%$) of the HI Rogues sample would have been selected by the VD05 color criteria for dry mergers. These 15 ``wet" Red Rogues (with neutral hydrogen in or around them) are displayed in Figure 3 with HI contours superimposed; the three normal ellipticals with peculiar HI are shown first, followed by the 12 peculiar ellipticals with HI. The presence of HI in these systems varies from lying in a well-defined disk and shells in Cen A (NGC 5128) to being confined in a gas-rich companion for NGC 4382. \\
\indent The optical morphologies of these Red Rogues appear very similar to the sample studied in VD05. 12 of the 15 Red Rogues are classified in the Rogues Gallery as pecEo and pecEi, as they exhibit signs of tidal interaction at low surface brightness levels in the form of tails and plumes of stars; the other three Red Rogues are EpecH, or normal ellipticals from the optical standpoint. This matches the behavior of van Dokkum's bulge-dominated population, 71\% of which -- the ``dry mergers" -- show tidal features, with the remainder exhibiting no clear sign of an interaction history down to very low surface brightness levels, making our ``wet" peculiar and ``wet" normal populations a counterpart to the ``red" peculiar and ``red" normal populations of VD05. A comparison of the deep co-added BVR optical imaging of cdfs-1100 from the MUSYC survey, utilized in VD05, to a DSS image of NGC 7135 (one of the Red Rogues) is shown in Figure 4; the galaxies in these two samples are clearly morphologically similar. \\
\indent That 15 out of 20 Rogues would have been selected by the VD05 dry merger color criteria is in itself an interesting result, considering that the Rogues are a sample selected to have gas. However, perhaps even more interesting is the ability to look for trends within the sample. In Figure 5, normal ellipticals and peculiar ellipticals are plotted using different symbols; no clear trend is apparent. This is possibly due either to our small sample size or to the fact that if we were to obtain deeper optical imaging (such as that used by VD05), it is possible that we would detect tidal features in our ``normal" early type systems. The two merger remnants (MRs), NGC 3921 and NGC 7252, are indicated by asterisks; they are clearly too blue to be selected by the VD05 criteria. In VD05, 18\% of the red objects are currently undergoing interactions, but our sample was chosen on the basis of looking almost exclusively at galaxies classified as ellipticals or peculiar ellipticals; we added two well known merger remnants for comparison, but clearly ongoing mergers are listed under other Rogues classes. We therefore selected against ``ongoing merger" systems in favor of systems which have likely already finished merging (i.e. the pecEo and pecEi galaxies, as well as the MRs). \\
\indent VD05 uses a subsample of merger pairs and merger remnants in order to estimate the effects that dry mergers would have on the evolution of the luminosity function of red galaxies. He finds that, with some caveats, dry mergers can explain the constant luminosity density out to z=1, which is incompatible with only passive evolution of red galaxies. We, however, find that if the ``wet" peculiar Red Rogues presented in this paper are actually the local analogues of these higher redshift merger remnants, then the VD05 systems are not necessarily dissipationless. \\

\subsection{Star Formation}

\indent It is interesting to note that one of the systems which falls unambiguously within the VD05 sample in Figure 2 is Markarian 315, which has a star formation rate of 30-40 M$\sol$ yr$^{-1}$ \citep{Ciroi05}. The presence of HI coupled with such a high rate of star formation makes this galaxy seem hardly dry. At least one other Red Rogue is also forming stars. Centaurus A (NGC 5128) has 1.5 $\times$ 10$^{8}$ M$\sol$ of HI associated with its shells and 4.5 $\times$ 10$^{8}$ M$\sol$ in the disk \citep{Schiminovich94}, where vigorous star formation is occurring \citep{Ebneter83}. CO has also been observed in both the disk (at the level of 2 $\times$ 10$^{8}$ M$\sol$; \citealt{Eckart90}), as well as in the shells, where it corresponds to 4.3 $\times$ 10$^{7}$ M$\sol$ of H$_{2}$ \citep{Charmandaris00}. Though Cen A has the appearance of simply an elliptical galaxy with a dust lane, it actually possesses a disk which is energetically forming stars despite its ``red" classification. The presence of gas does not necessarily imply star formation in the other Red Rogues, but in the cases of Mrk 315 and Cen A it does. \\
\indent In light of our findings, it is worth considering whether the evolutionary histories derived by VD05 are strongly constrained by the broadband filters available to him (BVRI). The inclusion of shorter and/or longer wavelength data (UV, $\it{U}$-band, NIR) or spectroscopy would better constrain these histories, especially by breaking the degeneracy between extinction and stellar ages (e.g. \citealt{GildePaz02}, \citealt{Anders04}) or by detecting ``frostings" of younger stars \citep{Trager01}. If they revealed a component of young stars, these better constrained star formation histories would then also constrain how ``dry" a previous merger could have been.

\section{Conclusion}

\indent In this paper, we have examined the selection criteria for picking out dry mergers as utilized in VD05 by comparing his sample to one consisting of early types with known HI properties. For our small sample of 20 systems, we find that 15 galaxies with HI (or 75\%) would have been selected by the VD05 dry merger criteria. VD05 invoked dry merging to explain the evolution of the luminosity function of bright red galaxies since z=1. However, we show that selecting dry mergers on the basis of their colors is not sufficient to ensure that the systems will not have gas. The presence of HI in turn allows for the possibility of significant star formation, and our sample contains at least two examples for which this is the case. \\

\acknowledgments
The authors wish to thank the referee, Dr. F. Schweizer, for his suggestions, which have greatly improved this manuscript. The authors also wish to thank A. Basu-Zych and B. Johnson for helpful comments and useful discussions. This work was supported in part by an NSF grant to Columbia University. We acknowledge the usage of the HyperLeda Database (http://leda.univ-lyon1.fr) and the NASA/IPAC Extragalactic Database (NED) which is operated by the Jet Propulsion Laboratory, California Institute of Technology, under contract with the National Aeronautics and Space Administration. Optical images were taken from the Digitized Sky Survey, produced at the Space Telescope Science Institute under U.S. Government grant NAG W-2166. The images of these surveys are based on photographic data obtained using the Oschin Schmidt Telescope on Palomar Mountain and the UK Schmidt Telescope.

\clearpage

\begin{table}[t]
\begin{center}
\centerline{Table~1: The Sample}
\medskip
\begin{tabular}{lcccccc}
\hline
\hline
Object & M$_{R}$ & B-R & d (Mpc) & R$_{ap}$ & Rogues Class. & Ref. \\
\hline
IC 2006 & -19.81 & 1.44 & 15.7 & 0.84 & EpecH & d \\ 
MCG -5-7-1 & -20.27 & 1.60 & 58.3* & 0.98 & pecEi & i \\
Mrk 315 & -22.11 & 1.65 & 157* & 1.01 & pecEi & j \\
NGC 474 & -20.92 & 1.54 & 32.5 & 1.05 & pecEo & c* \\ 
NGC 680 & -21.25 & 1.68 & 37.7* & 0.95 & pecEi & * \\
NGC 1052 & -21.27 & 1.62 & 17.8 & 1.04 & EpecH & g \\
NGC 1210 & -20.33 & 1.25 & 48.7* & 1.04 & pecEi & d \\
NGC 1316 & -22.25 & 1.55 & 16.9 & 0.97 & pecEo & a \\ 
NGC 2768 & -21.69 & 1.44 & 23.7 & 0.96 & EpecH & g \\
NGC 2865 & -21.24 & 1.43 & 35.7 & 0.84 & pecEi & e \\
NGC 3921 & -21.77 & 1.27 & 81.7* & 1.26 & MR & b \\
NGC 4125 & -21.96 & 1.58 & 24.2 & 1.10 & pecEo & c \\
NGC 4382 & -21.67 & 1.45 & 16.8 & 0.90 & pecEo & h \\
NGC 4406 & -21.60 & 1.49 & 16.8 & 1.09 & pecEo & c \\
NGC 5018 & -22.12 & 1.53 & 40.9 & 0.96 & pecEo & e \\
NGC 5128 & -21.07 & 1.65 & 3.8** & 1.12 & pecEi & d \\ 
NGC 5903 & -21.26 & 1.71 & 35.9 & 1.05 & EpecH & a \\
NGC 7135 & -20.75 & 1.64 & 34.7 & 1.11 & pecEo & c \\
NGC 7252 & -21.78 & 1.31 & 62.9* & 1.01 & MR & c \\
NGC 7626 & -21.42 & 1.72 & 46.0* & 0.87 & pecEo & f \\
\hline
\end{tabular}
\end{center}
\caption{\small The sample. Columns display object name, 
M$_{R}$ in the Cousins system, B-R, distance (Mpc), aperture fraction (defined in the text), HI classification in the Rogues Gallery \citep{RoguesGallery}, and references to the photometry: a, \citealt{Sandage75}; b, \citealt{Huchra77}; c, \citealt{Sandage78}; d, \citealt{Lauberts89}; e, \citealt{Poulain94}; f, \citealt{Sandage73}; g, \citealt{Peletier90}; h, \citealt{Schroeder96}; i, \citealt{Lauberts84}; j, \citealt{Moles87}. Distances are taken from \citet{Tully88}, except for those marked with an asterisk (*), which are calculated using Virgo-corrected velocities from NED and assuming H$_{o}$=75 \kms Mpc$^{-1}$ in order to be consistent with \citet{Tully88}. The distance to NGC 5128 is taken from \citet{Rejkuba05}. Photometry from references a, b, c, f, and k have been converted to Cousins. The photometry for NGC 474 is marked as suspect by \citet{Sandage78}, and the unpublished photometry for NGC 680 is available through the HyperLeda aperture photometry database (a description of this specific database is available in \citet{Prugniel98}). A Rogues classification of pecEo refers to peculiar ellipticals with HI outside of the optical body, pecEi refers to peculiar ellipticals with HI inside of the optical body, EpecH refers to normal early type galaxies with peculiar HI, and MR refers to merger remnants.}
\end{table}

\clearpage

\begin{figure}
\epsscale{1.0}\plotone{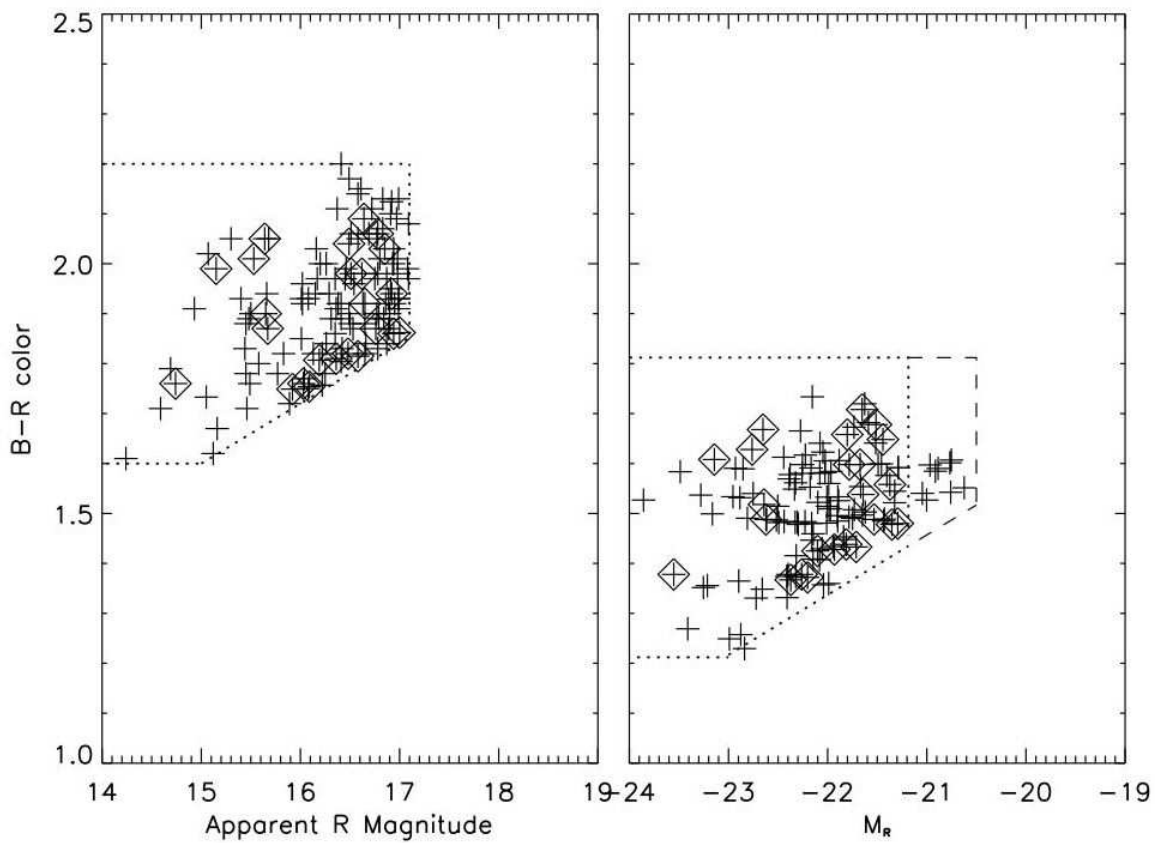}
\caption{($\it{left}$) CMD of sample used in VD05 at z=0.1; R magnitudes are apparent. ($\it{right}$) Same sample adjusted to z=0 and k-corrected; R magnitudes are absolute. Adjusting the z=0.1 sample (at left) to z=0 (at right) makes the specified apparent-magnitude dependent criterion no longer a valid photometric constraint on the population. All crosses encircled with diamonds represent galaxies for which the redshift is not known; for these galaxies, we have assumed z=0.1.}
\end{figure}

\begin{figure}
\epsscale{1.0}\plotone{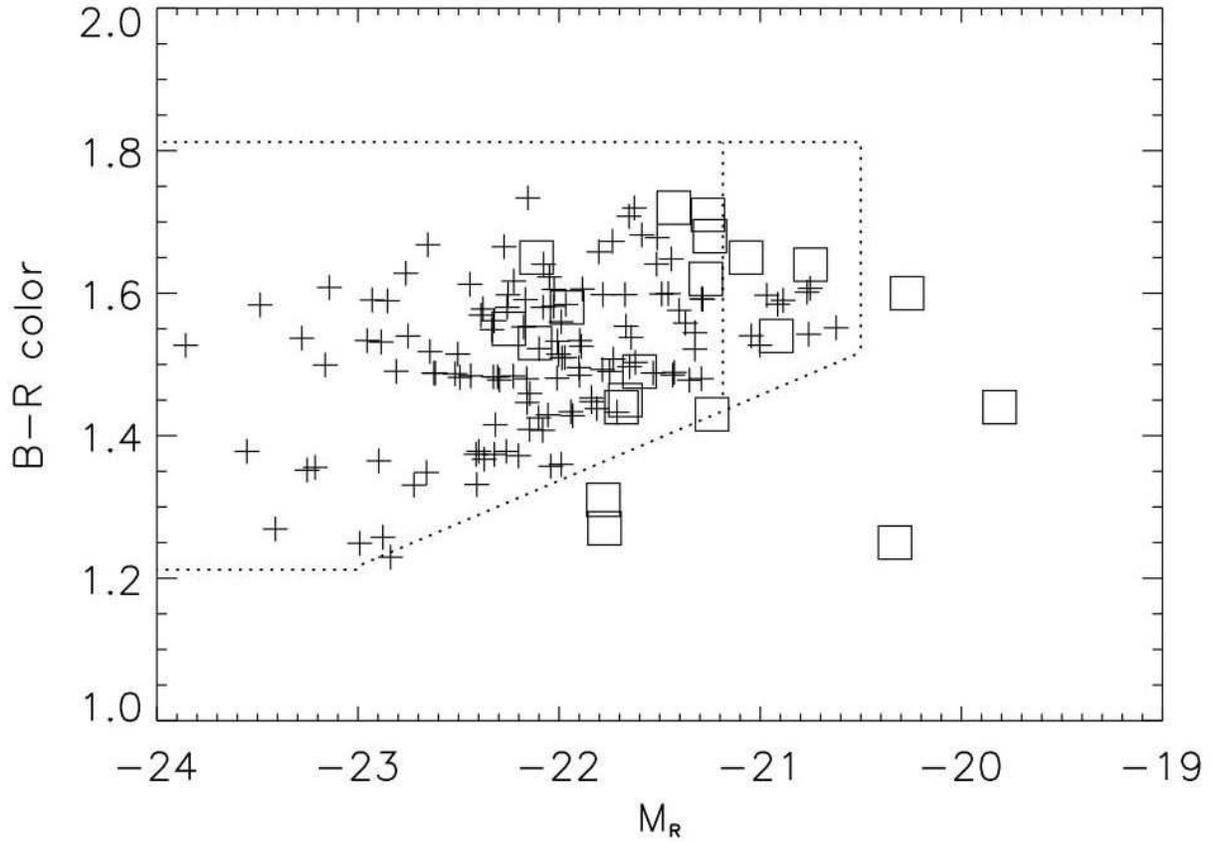}
\caption{CMDs of the HI sample (squares) and the VD05 sample (+) with both data sets presented in M$_{R}$. The dotted lines display the VD05 selection criteria as well as our extension of the color and magnitude criteria to fainter magnitudes.}
\end{figure}

\begin{figure}
\begin{tabular}{ccc}
\includegraphics[height=4.35cm]{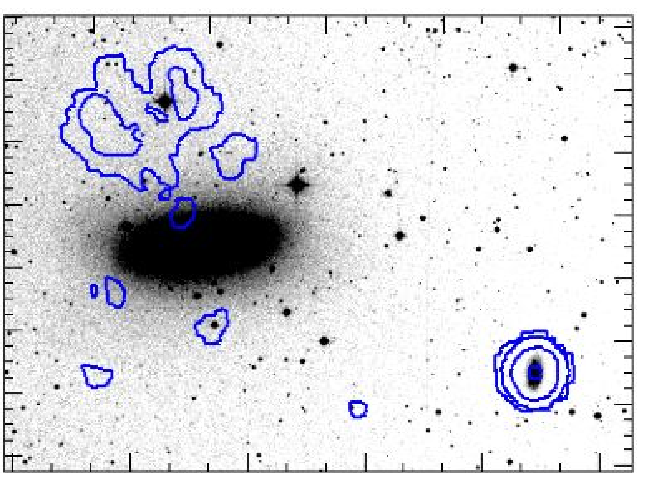}
\includegraphics[height=4.35cm]{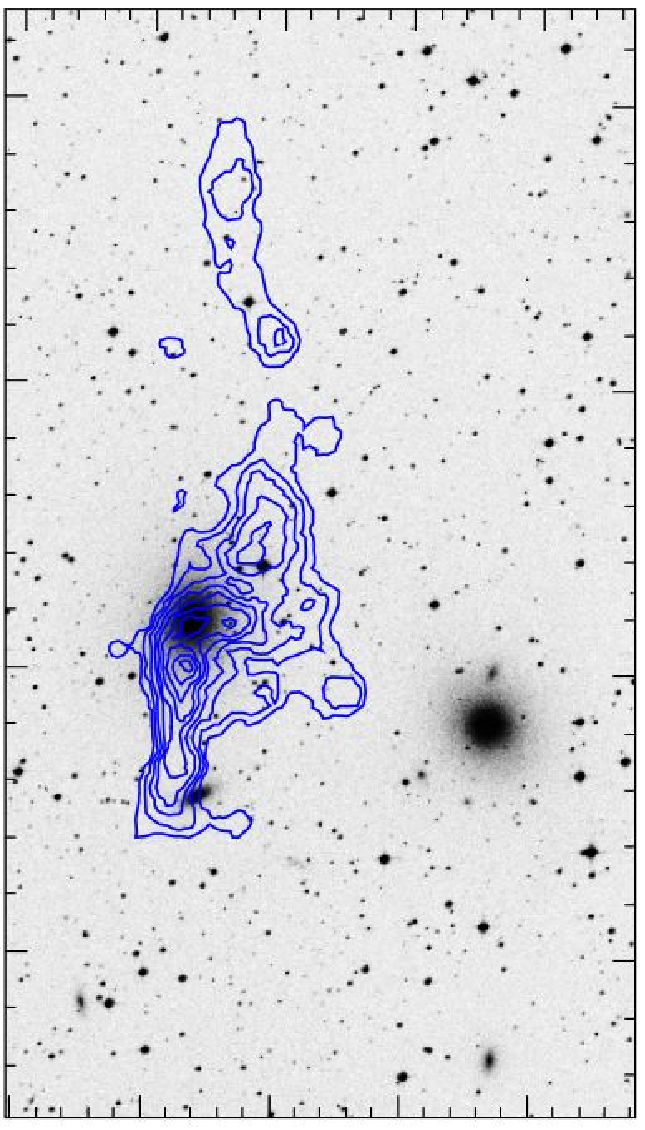}
\includegraphics[height=4.35cm]{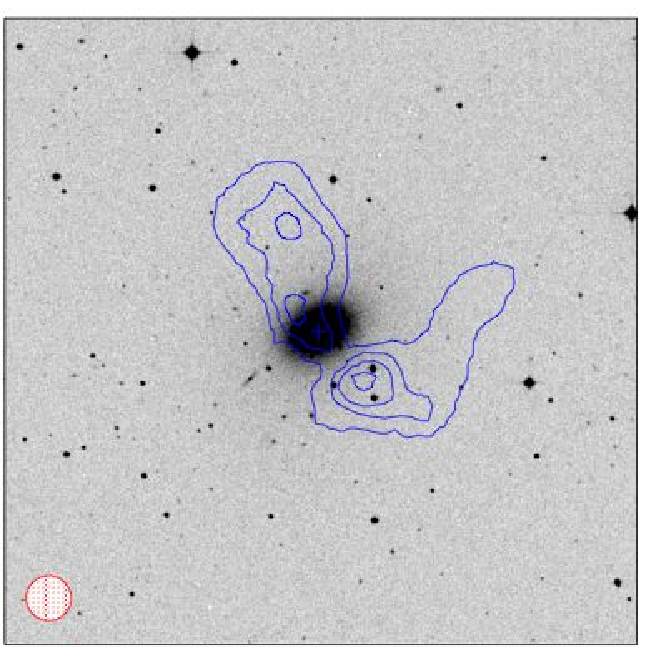} 
\includegraphics[height=4.35cm]{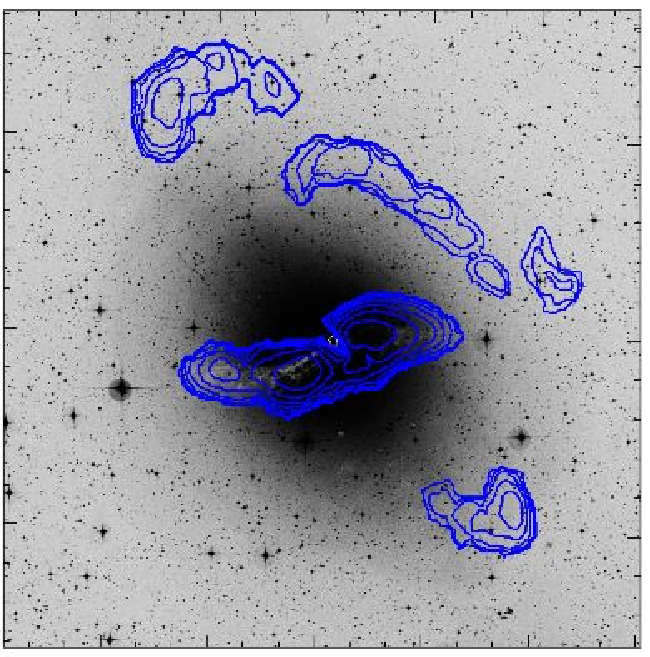} \\

\includegraphics[height=3.9cm]{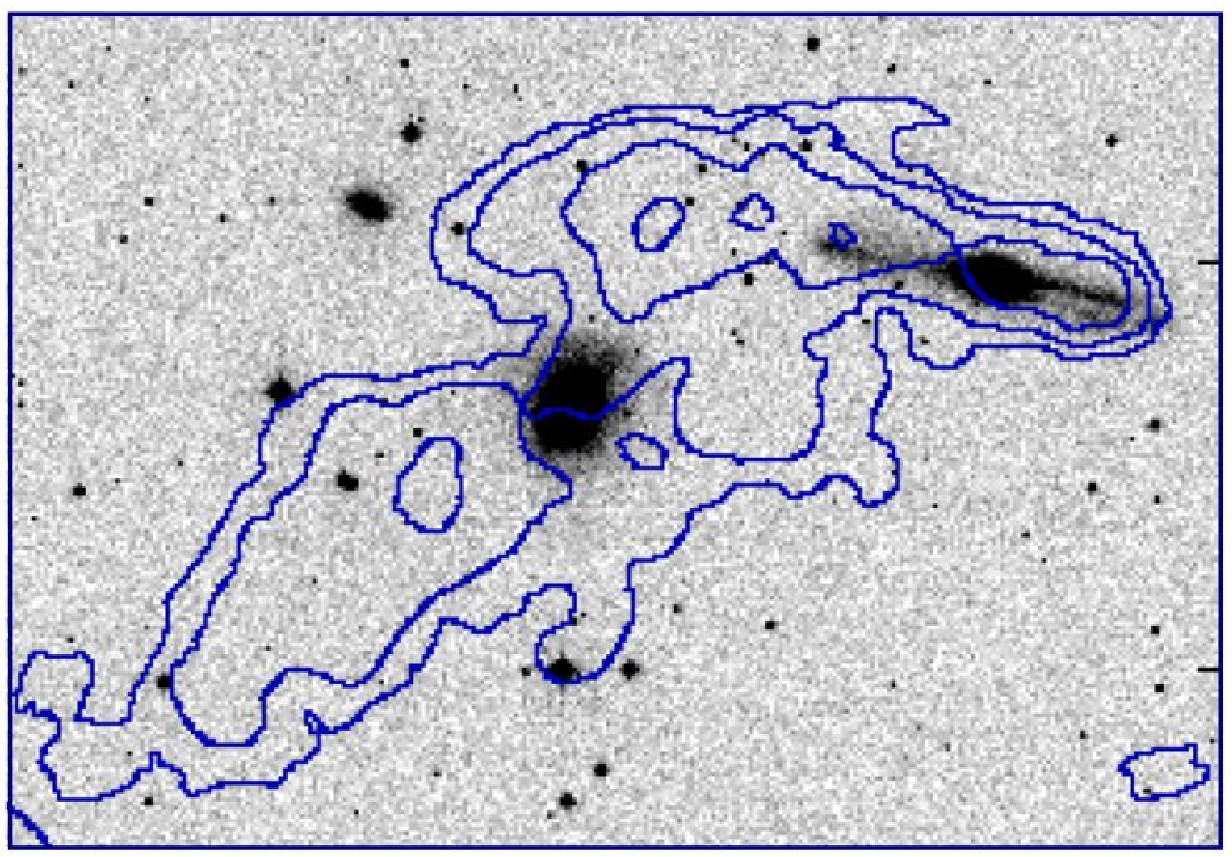}
\includegraphics[height=3.9cm]{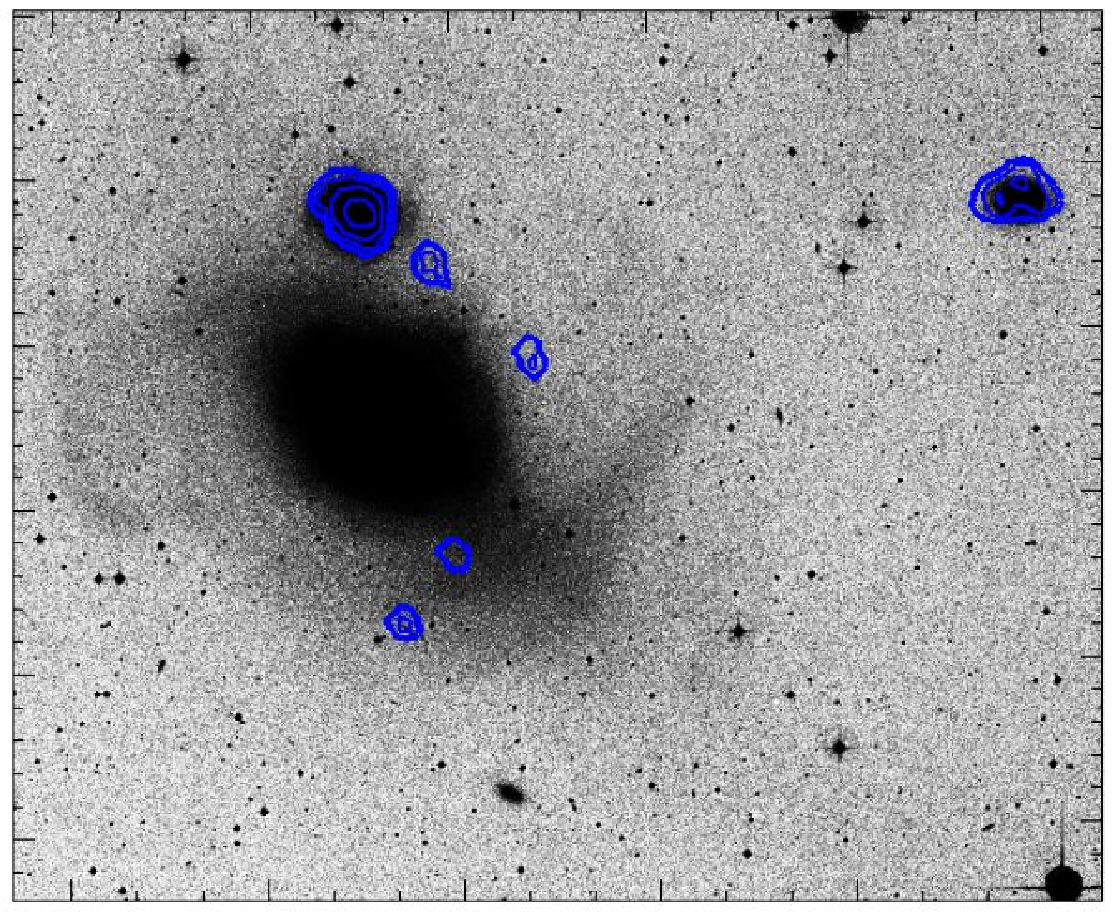}
\includegraphics[height=3.9cm]{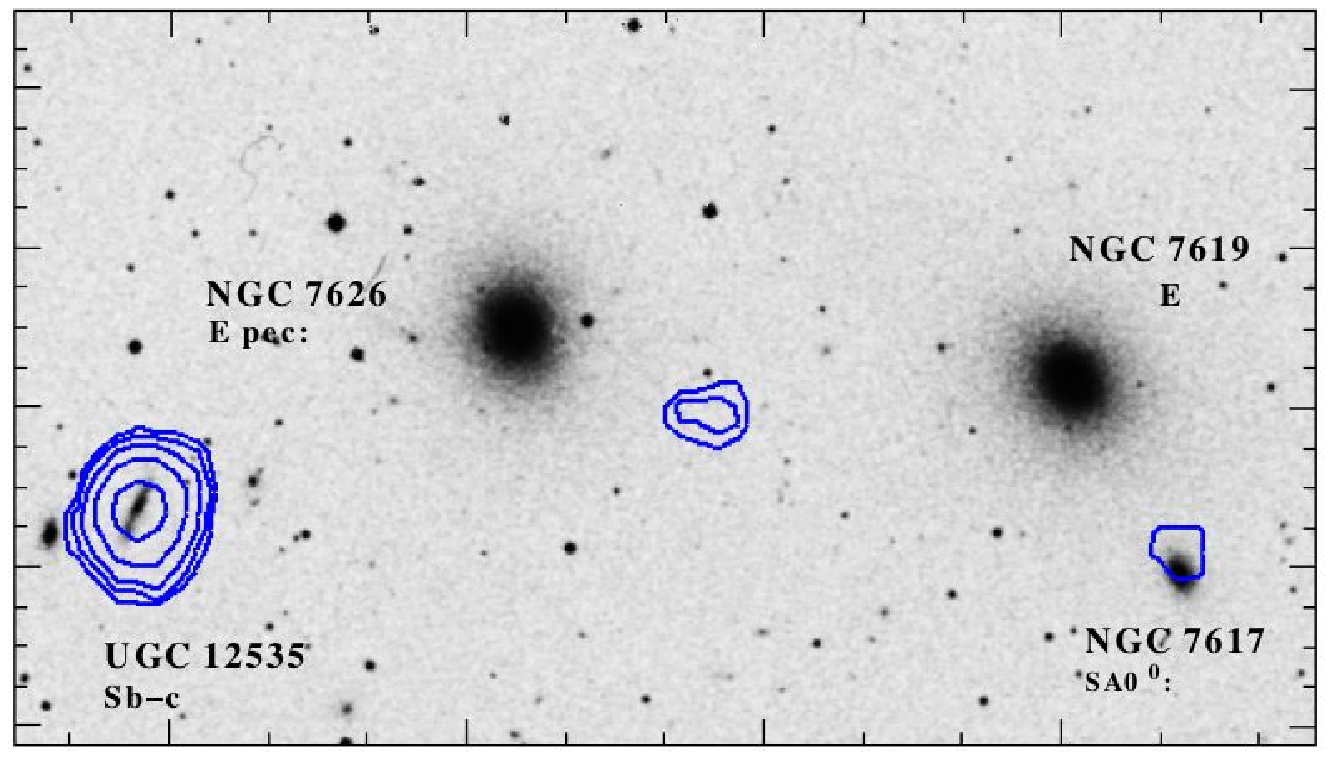} \\

\includegraphics[height=3.75cm]{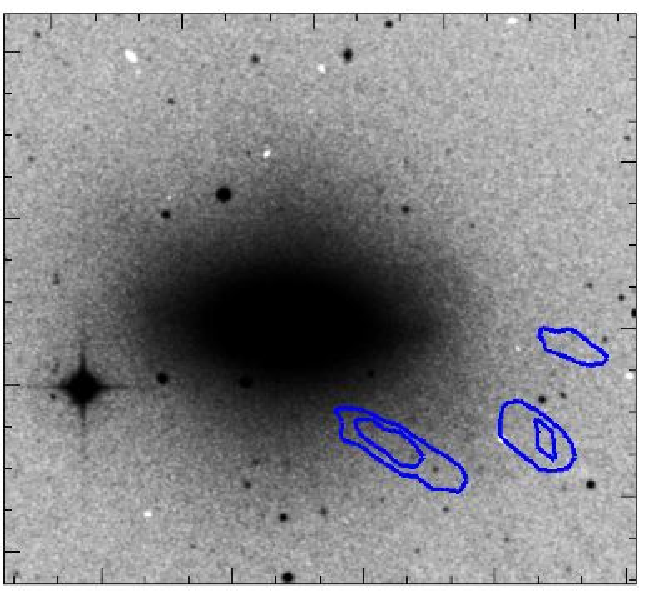}
\includegraphics[height=3.75cm]{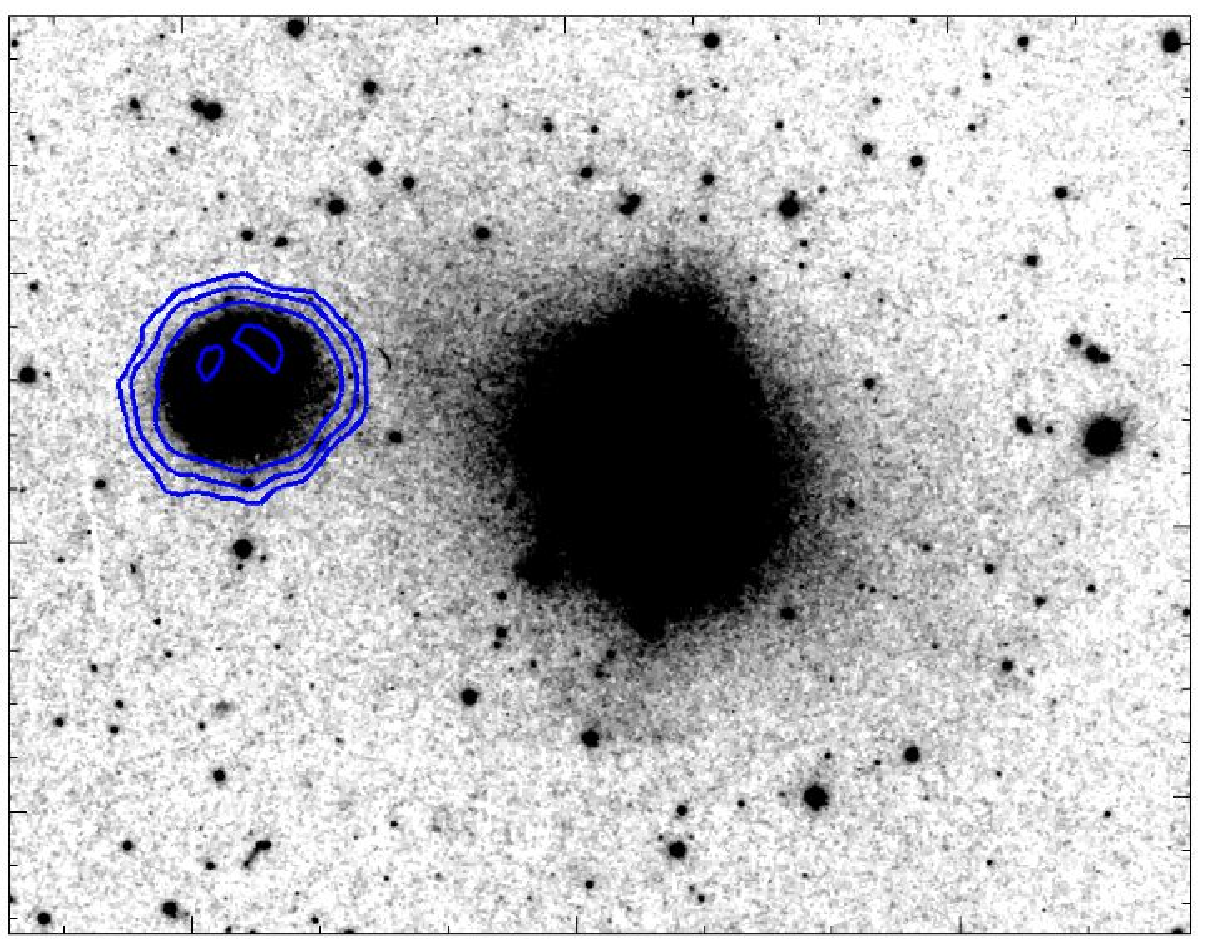}
\includegraphics[height=3.75cm]{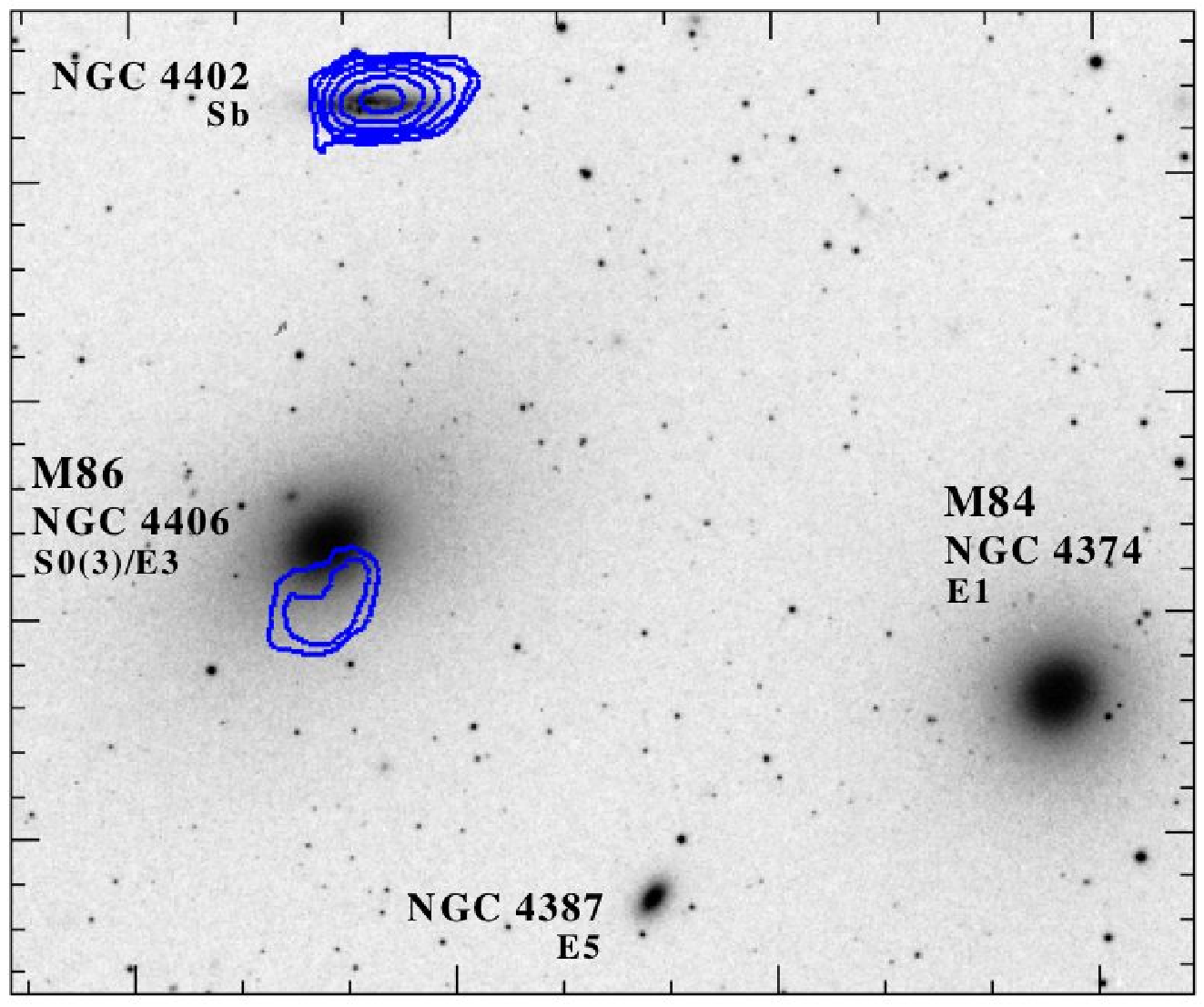}
\includegraphics[height=3.75cm]{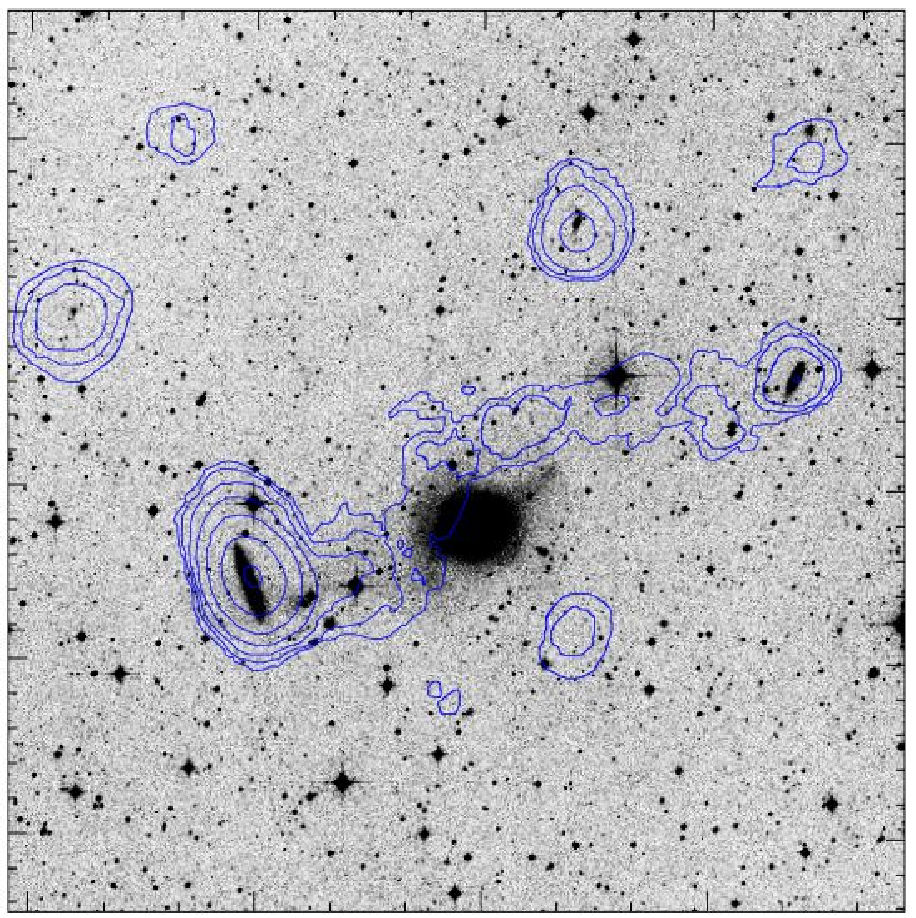} \\

\includegraphics[height=4.7cm]{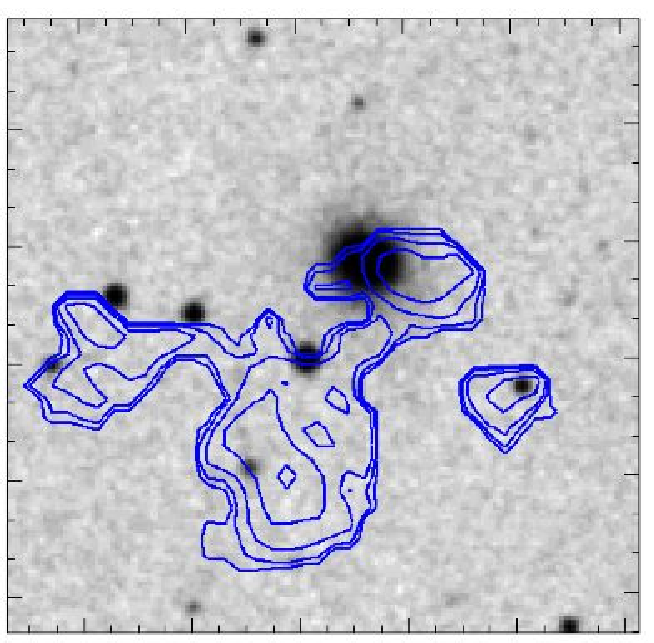}
\includegraphics[height=4.7cm]{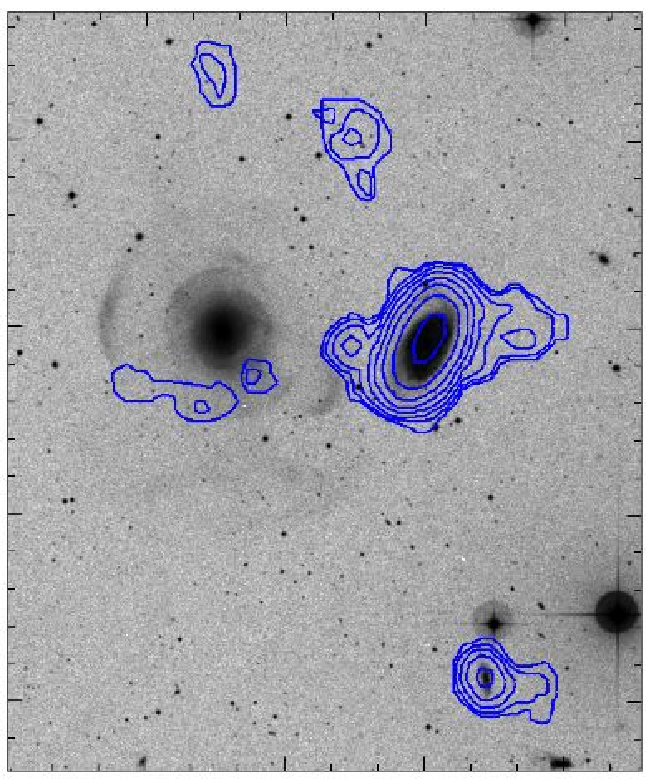}
\includegraphics[height=4.7cm]{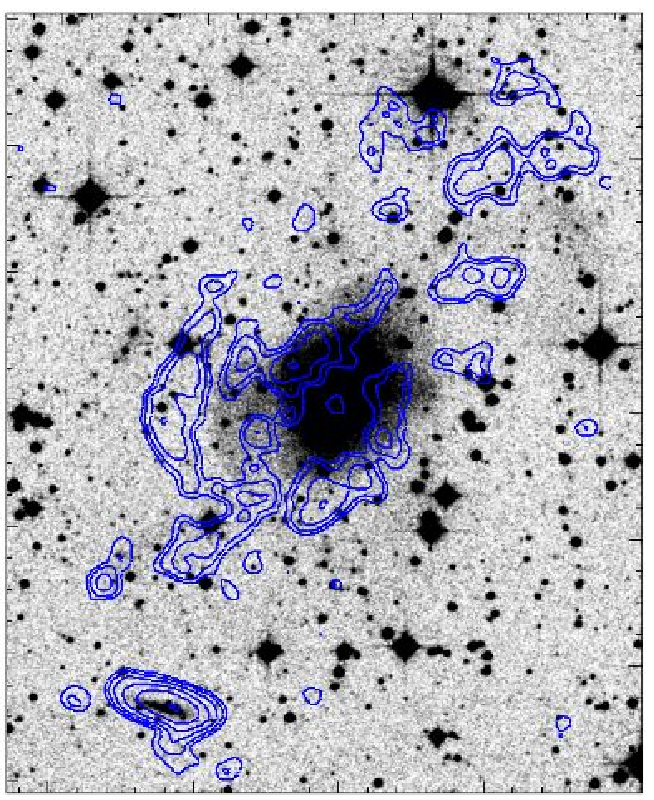}
\includegraphics[height=4.7cm]{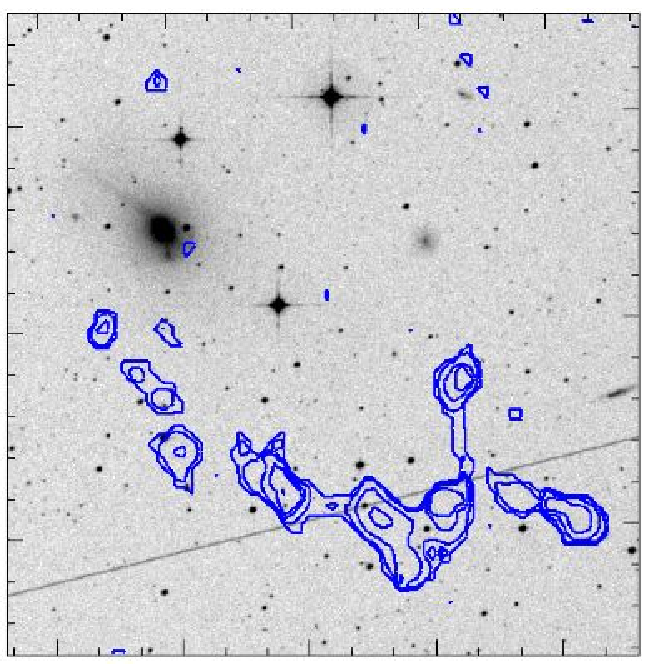} \\

\end{tabular}
\caption{The Red Rogues, as published in the Rogues Gallery, with HI contours overlaid on optical DSS images. Row 1: NGC 2768 \citep{Schiminovichetal01}, NGC 5903 \citep{Appleton90}, NGC 1052 \citep{vanGorkom86}, NGC 5128 \citep{Schiminovich94}; Row 2: NGC 680 \citep{vanMoorsel88}, NGC 1316 \citep{Horellou01}, NGC 7626 \citep{Hibbard01}; Row 3: NGC 4125 \citep{Rupen01}, NGC 4382 \citep{Hibbard01}, NGC 4406 \citep{Li01}, NGC 5018 \citep{Kim88}; Row 4: Mrk 315 \citep{Simkin01}, NGC 474 \citep{Schiminovichetal01}, NGC 2865 \citep{Schiminovich95}, NGC 7135 \citep{Schiminovichetal01}.}

\end{figure}

\begin{figure}
\plotone{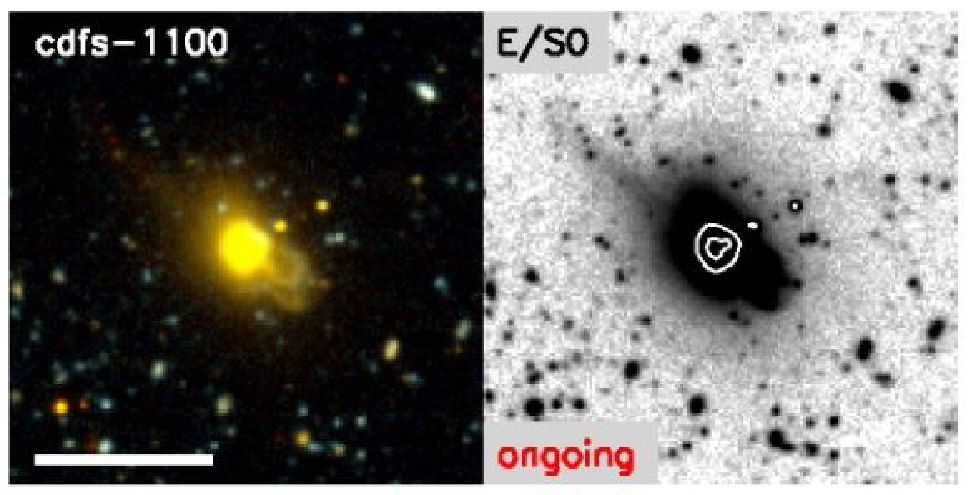}
\plotone{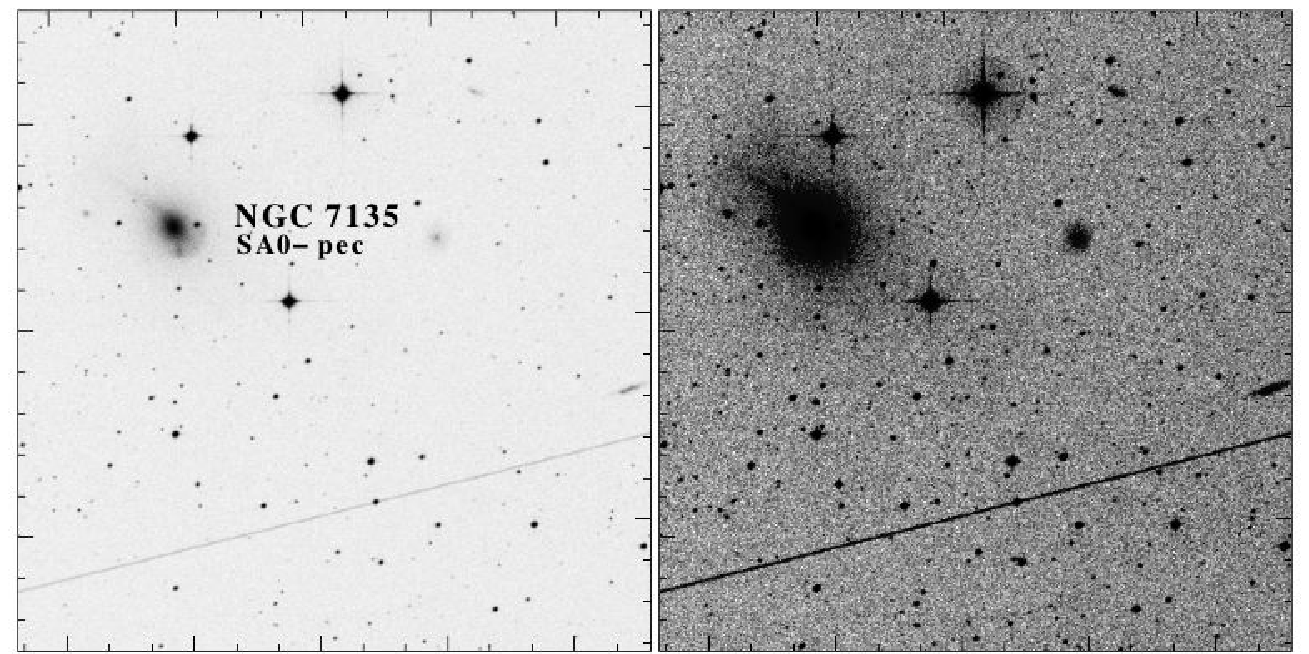}
\caption{Deep optical imaging of two morphologically similar galaxies: cdfs-1100 from VD05 (top) and NGC 7135 from the Red Rogues sample (bottom).}
\end{figure}

\begin{figure}
\epsscale{1.0}\plotone{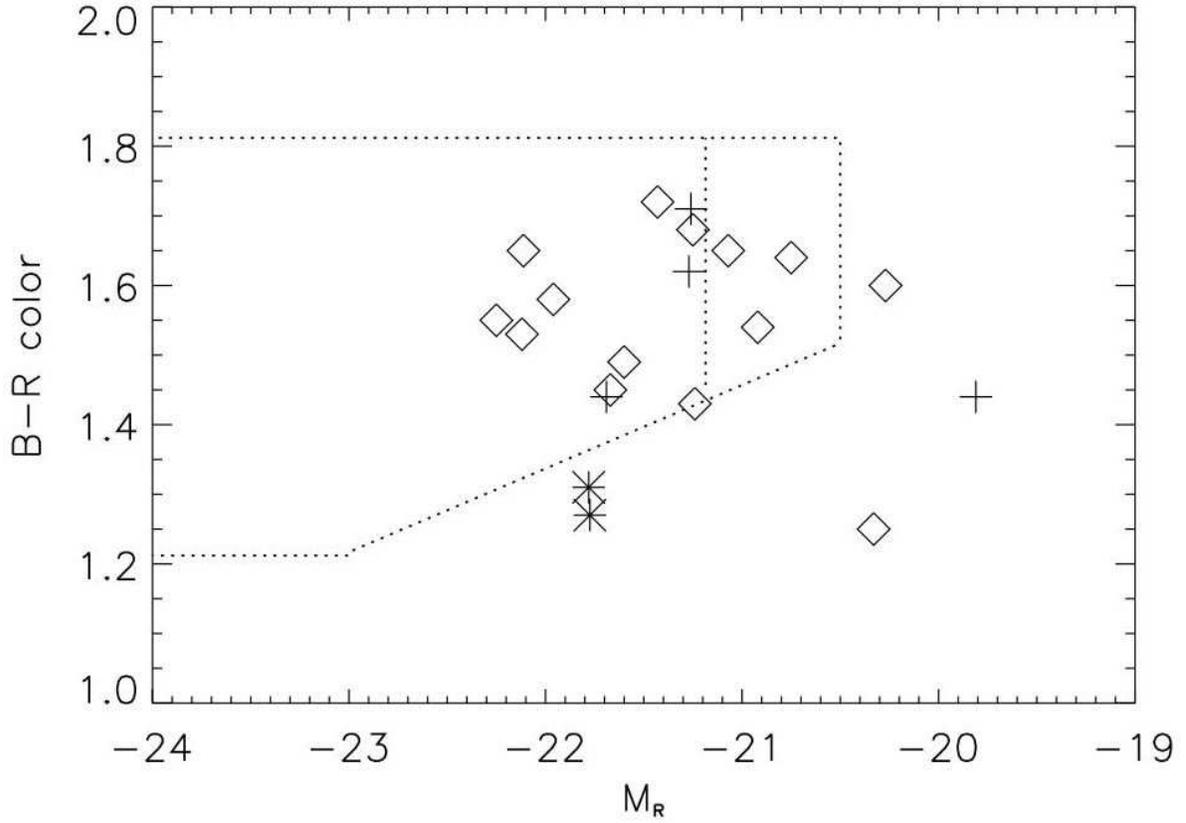}
\caption{HI sample divided into normal (+) and peculiar (diamonds) ellipticals. The color and magnitude selection criteria used in this paper are also displayed. The two asterisks (*) indicate the merger remnants, which are two of the bluest systems in the HI sample.}
\end{figure}

\clearpage


\end{document}